\documentclass[12pt]{iopart}
\usepackage{graphicx}
\usepackage{xfrac}
\usepackage{bm}        
\usepackage{amssymb}   
\usepackage{color}
\usepackage{subfig}
\usepackage{mathrsfs}
\usepackage{latexsym}

\begin{document}

\title[Heading]{Viscosity Bound Violation in Viscoelastic Fermi Liquids}

\author{M P Gochan$^1$, Hua Li$^1$, K S Bedell$^1$}

\address{$^1$ Department of Physics, Boston College, Chestnut Hill, MA 02467, United States of America }
\ead{gochan@bc.edu}
\vspace{10pt}
\begin{indented}
\item[]February 2019
\end{indented}

\begin{abstract}
The anti-de Sitter/conformal field theory correspondence (AdS/CFT) has been used to determine a lower bound on the ratio of shear viscosity $\left(\eta\right)$ to entropy density $(s)$ for strongly-coupled field theories with a gravity dual. The conjectured universal lower bound, given as $\eta/s\geq\hbar/4\pi k_B$, is a measure of interaction strength in a quantum fluid where equality indicates a perfect quantum fluid. In this paper we study $\eta/s$ in a Fermi gas in the unitary limit. We show that in addition to a local minimum for $\eta/s$ at $T\approx 2T_c$ which obeys the lower bound, a more interesting result exists in the violation of the $\eta/s$ lower bound due to the superfluid fluctuations above $T_c$. To conclude, we examine the viscoelastic properties of the unitary Fermi gas. Previous work brought to light the connection between violation of the $\eta/s$ bound and a viscoelastic response in the context of holographic solids. We ultimately find that, in addition to holographic solids, all Fermi liquids with a viscoelastic response produced by superfluid fluctuations can violate the universal $\eta/s$ lower bound.
\end{abstract}

%
%
%
%
%

\section{Introduction}
Using the anti-de Sitter/conformal field theory (AdS/CFT) correspondence, strongly interacting quantum field theories can be described in terms of weakly interacting gravitational systems. This has led to the conjecture that there exists a lower bound ---the KSS bound--- for $\eta/s$ in a strongly coupled field theory given by \cite{Cremonini_2011,Kovtun_Son_Starinets_2005,KSS,Iqbal,YangMillsPlasma}
\begin{equation}
\frac{\eta}{s} \geq \frac{\hbar}{4\pi k_{\scriptscriptstyle{\text{B}}}}
\label{KSS}
\end{equation}
Quantum fluids of varying density, such as the quark gluon plasma and the unitary Fermi gas, that obey eqn. (\ref{KSS}), are called nearly perfect quantum liquids where equality denotes a perfect quantum liquid \cite{NPF,Thomas_2010}. The AdS/CFT correspondence additionally creates a bridge between gravitational physics and condensed matter physics and allows one to be studied in terms of the other \cite{Phonon1,Phonon2}. It's been shown that as the unitary Fermi gas undergoes a superfluid phase transition, superfluid fluctuations above the transition temperature, $T_c$, have significant effects on the spin transport \cite{Sommer_Zwierlein_2011,Li_Jackiewicz_Bedell_2015}. This result is the motivation for our work. We sought to determine if such superfluid fluctuations could have a similar impact in viscosity and subsequently the KSS bound. Recent experiments on the unitary Fermi gas $^6\text{Li}$ show a normal/superfluid phase transition at a transition temperature, $T_c \approx 0.167T_F$ \cite{Zwierlein}, where $T_F$ stands for the Fermi temperature. As for the viscosity, recent advances in experiments have allowed for its measurement \cite{Cao_Thomas_2011} and subsequently led to the measurement of the ratio. Such measurements show a minimum that obeys the bound given by (\ref{KSS}) at temperatures $T\approx 2T_c$ \cite{Schafer,Turlapov,Joseph_Thomas_VS}.

To better understand $\eta/s$ within the context of strongly correlated systems, we develop a simple theoretical model to calculate the quasiparticle scattering rates of a strongly correlated quantum liquid above $T_c$. Such a model differs from past calculations \cite{PRA,SMatrix,Salasnich,Pakhira} in that we include the effects of superfluid fluctuations as $T\rightarrow T_c^+$. The model separates the quasiparticle scattering amplitude for the strongly correlated quantum fluid into two components: the superfluid fluctuations term coming from the particle-particle pairing fluctuations in the singlet scattering channel above $T_c$, and a normal Fermi liquid scattering term calculated from the local version of the induced interaction model \cite{Engelbrecht_Bedell_1995,Jackiewicz_Bedell_2005}. Applying our theory to the unitary Fermi gas, we calculate $\eta/s$ for the unitary Fermi gas about $T_c$ following the methods used in the transport studies of Landau Fermi-liquid theory \cite{Baym_Pethick}. We find a local minimum as $T\rightarrow T_c$ of $\eta/s \approx 0.3\hbar/k_{\scriptscriptstyle{\text{B}}}$ which agrees with the experimentally measured lower bound \cite{Cao_Thomas_2011}. However, an additional intriguing result of $\eta/s$ dropping to zero at $T_c$ thus violating (\ref{KSS}). Our work therefore seeks to explain the nature of this violation within the context of Landau Fermi liquid theory. While violations of (\ref{KSS}) are not uncommon, for example the work done by Alberte et. al. \cite{Visco2} and Jain et. al. \cite{Jain} both show violation, our work is unique in that our calculation is done for the unitary Fermi gas, a system frequently studied experimentally. Furthermore, we differ from other work on $\eta/s$ in the unitary Fermi gas, such as that by Samanta et. al. that also showed violation \cite{Samanta}, in that the system under consideration was trapped. While our result appears to be in contradiction with the work done by Cao et. al. \cite{Cao_Thomas_2011}, we find good qualitative agreement with the more recent analysis done by Joseph et. al. \cite{Joseph_Thomas_VS}. We believe this discrepancy is because the measurements were done over a wide temperature range while the violation of the bound happens in a small window around $T_c$. Additionally, due to the breakdown of the quasiparticle picture, numerous other methods have been employed such as those performed by Enss et. al. \cite{Enss} to determine the viscosity. While we don't disagree with these results, we feel our model is valid due to the experimental support of the quasiparticle picture near $T_c$ (as shown in Fig. 4 and will be discussed later). To conclude, we draw on previous work by  Alberte, Baggioli, and Pujol\`as \cite{Visco2,Baggioli} they present the idea of the viscoelastic nature of holographic solids violating the bound. We expand on their work and provide insight into this high-energy problem from the viewpoint of condensed matter. 

\section{Superfluid Fluctuations in the Unitary Fermi Gas}
The high transition temperature, $T_c \approx 0.167 T_F$,  of the unitary Fermi gas allows for the experimental measurement of $\eta/s$ at temperatures close to $T_c$ , where superfluid fluctuations could play a role \cite{Zwierlein,Cao_Thomas_2011}. For example, previous study of spin transport found that superfluid fluctuations play a significant role in the spin diffusion \cite{Sommer_Zwierlein_2011,Li_Jackiewicz_Bedell_2015}. As such, our work sets out to understand how the superfluid fluctuations may affect the viscosity and subsequently $\eta/s$. The superfluid fluctuations come from the particle-particle pairing fluctuations in the spin singlet quasiparticle scattering channel closely above $T_c$. 
Due to the pairing fluctuations, the quasiparticle scattering amplitudes for small total momentum scattering diverge at $T_c$. Here we consider only the s-wave (spin singlet) pairing mechanism for the Cooper pairs and incorporate the superfluid fluctuations in the scattering amplitudes by evaluating the temperature vertex function of particle-particle type in the spin singlet channel for small total momentum scattering using standard quantum field theory methods \cite{Abrikosov}. The spin singlet temperature vertex function $\mathscr{T}_s(K)$ is generated from the diagram shown in Fig. {\ref{fig:diagram2}}, leading to the following integral equation:
\begin{figure}
	\centering
	\includegraphics[scale=0.4]{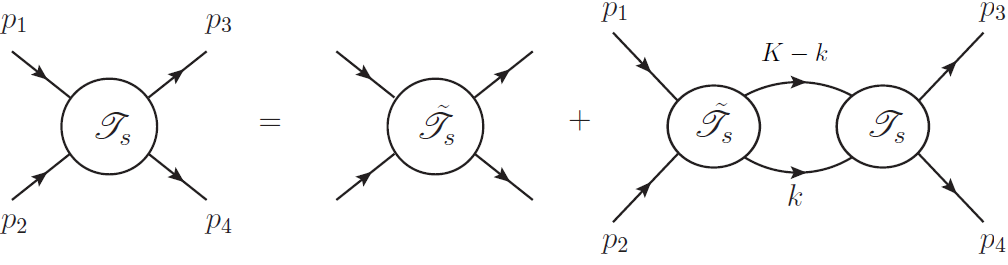}
	\caption{Feynman diagram for the temperature vertex function of particle-particle type, $\mathscr{T}_s$. The bubbles represent the irreducible ($\tilde{\mathscr{T}}_s$) and fully reducible ($\mathscr{T}_s$) particle-particle vertex function, the solid lines stand for the fermion Green's functions. The propagators are the quasiparticle propagators with fully renormalized quasiparticle interactions and $K=p_1+p_2$.}
	\label{fig:diagram2}
\end{figure}
\begin{eqnarray}
\mathscr{T}_s(p_1,p_2;p_3,p_4)&=& \tilde{\mathscr{T}}_s(p_1,p_2;p_3,p_4)-\frac{T}{2(2\pi)^3} \nonumber \\ && \times\Sigma_{\omega_n}\int\tilde{\mathscr{T}}_s(p_1,p_2;k,K-k)
\mathcal{G}(K-k) \nonumber \\ && \times \: \mathcal{G}(k)\mathscr{T}_s(k,K-k;p_3,p_4)\mathrm{d}^3k
\label{eq:Integral_Equation}
\end{eqnarray}
where, $p_i=(\mathbf{p}_i,\omega_i)$ are the four momenta of the scattering particles, and, $K=(\mathbf{K},\omega_0)$ stands for the total momentum of the incident particles. $\mathscr{T}_s$ depends only on the total momentum $K$,  $\mathscr{T}_s(p_1,p_2;p_3,p_4)\equiv\mathscr{T}_s(K)$, when $|\mathbf{p}_i|=k_F$ for $i=1,\cdots,4$ and $|\mathbf{K}|\ll k_F$. Solving eqn. (\ref{eq:Integral_Equation}), we can express $\mathscr{T}_s$ in the small $K$ limit as
\begin{equation}
\mathscr{T}_s(\mathbf{K},0)=\frac{1}{\frac{mp_f}{4\pi^2} \left[\ln\frac{T}{T_c}-\frac{1}{6}\left(\frac{v_f|\mathbf{K}|}{2\omega_D}\right)^2-\frac{7\zeta(3)}{3\pi^2} \left(\frac{v_f|\mathbf{K}|}{4T}\right)^2\right]}
\label{eq:Vertex_Function_Result}
\end{equation}
where $T_c=\frac{2\gamma\omega_D}{\pi}e^{-4\pi^2/mp_f|\tilde{\mathscr{T}}_s|}$, $\gamma$ is the Euler-Mascheroni constant, $p_f$ is the Fermi momentum, and $\omega_D=0.244\varepsilon_F$ is the cutoff frequency \cite{Gorkov_Barkhudarov_1961}. $\tilde{\mathscr{T}}_s$ is the zero temperature irreducible particle-particle vertex function, which is approximately equal to the spin singlet normal Fermi-liquid scattering amplitude, denoted by \textit{a}, given diagramatically in Fig. \ref{fig:diagram}b \cite{Baym_Pethick}. In order to calculate the viscosity of the unitary Fermi gas, we need the normal Fermi-liquid scattering amplitude. The total quasiparticle scattering probability, $\langle W\rangle\equiv\int\frac{\mathrm{d}\Omega}{4\pi}\frac{W(\theta,\phi)}{\cos(\theta/2)}$, is obtained by averaging the quasiparticle scattering amplitudes of different $\textbf{K}$'s over the phase space \cite{Baym_Pethick}. For the unitary Fermi gas, $\langle W\rangle$ is separated into a superfluid fluctuations term, $\langle W\rangle_{\rm{fluctuations}}$, and a normal Fermi-liquid scattering term, $\langle W\rangle_{\rm{normal}}$:
\begin{eqnarray}
\langle W\rangle & = & \int_0^{\mathbf{K}_{\text{max}}}\frac{\mathrm{d}\Omega}{4\pi}\frac{W_{\text{f}}(\theta,\phi)}{\cos(\theta/2)}
+\int_{\mathbf{q}_{\text{max}}}^{2P_f}\frac{\mathrm{d}\Omega}{4\pi}\frac{W_{\text{n}}(\theta,\phi)}{\cos(\theta/2)}\nonumber \\
& = & \langle W\rangle_{\rm{fluctuations}}+\langle W\rangle_{\rm{normal}}
\label{eq:Total_Scattering_Probability}
\end{eqnarray}
$\mathbf{K}_{max}$ stands for the critical value of the total momentum of the incident particles, beyond which Cooper pairs start to break down and the particles scatter off of each other as in the normal Fermi liquid state. It is given by $v_F|\mathbf{K}_{max}|=6\varpi$, where $\varpi=2\omega_D e^{-4\pi^2/mp_f |\tilde{\mathscr{T}}_s|}$, from regular quantum field theory analysis \cite{Abrikosov}. It's important to note that the angular averages in eqn. (\ref{eq:Total_Scattering_Probability}) are different due to the different angular dependencies in $\mathbf{K}_{max}$ and $\mathbf{q}_{max}$ \cite{Li_Jackiewicz_Bedell_2015,Ainsworth_Bedell_1987}.

The Landau parameters needed for computing the quasiparticle scattering amplitudes are determined from the local induced interaction model, shown diagramtically in Fig.\ref{fig:diagram}. First developed to study the quasiparticle interactions in liquid $^3$He, it has seen success in applications to other interacting Fermi systems and been further generalized to account for the momentum dependence in the scattering amplitudes \cite{Ainsworth_Bedell_1987,Babu_1973,ABBQ_1983,Bedell_Quader_1985}. According to the model, the quasiparticle interaction parameter, $f$, is generated from a direct term, $d$, which is equivalent to a model dependent effective quasiparticle potential, and an induced term coming from the coupling of collective excitations to the quasiparticles. The mechanism is shown diagrammatically in Fig.1 in Li et. al. \cite{Li_Jackiewicz_Bedell_2015}
\begin{figure}[h]
	\centering
	\includegraphics[scale=0.75]{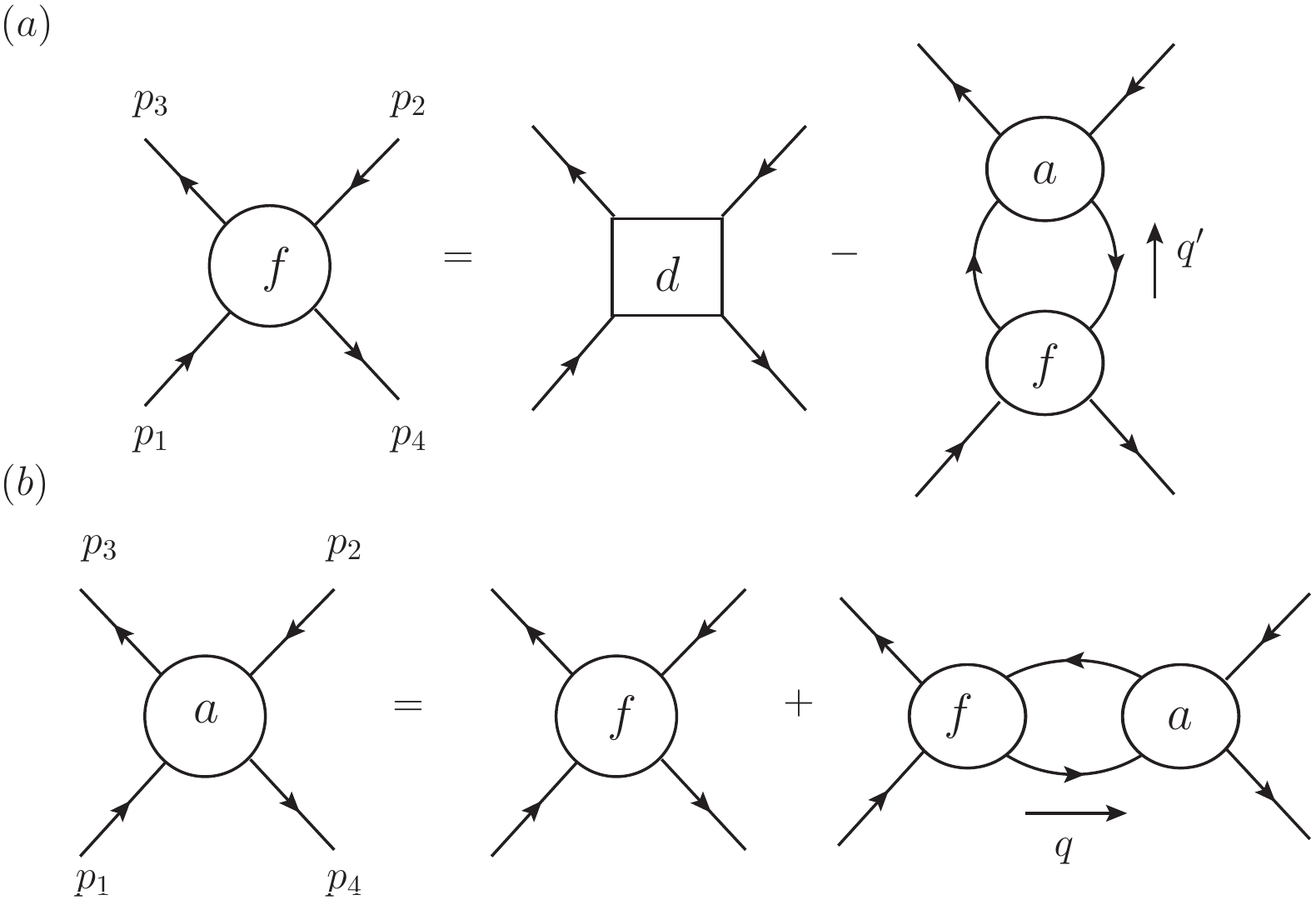}
	\caption{Diagrammatic representation of the induced interaction model. (a) represents the equation for Landau parameters $f$ decomposed into direct and induced terms; (b) sums all the reducible diagrams. It represents the equation relating $f$ to the scattering amplitudes$a=A/N(0)$. The momentum in the particle-hole channel is represented by $q=p_1-p_3=p_4-p_2$ and the momentum in the exchange particle-hole channel $q'=p_1-p_4=p_3-p_2$.}
	\label{fig:diagram}
\end{figure}
In this work we use a local, momentum independent, version of the induced interaction model where only the $l=0$ Landau parameters, $F_0^{s,a}$, are nonzero \cite{Li_Jackiewicz_Bedell_2015,Engelbrecht_Bedell_1995,Jackiewicz_Bedell_2005,Gaudio_Bedell_2009} and given as
\begin{equation}
F_0^s = D_0^s+\frac{1}{2}F_0^sA_0^s+\frac{3}{2}F_0^aA_0^a,
\label{eq:Landau_Parameters_s}
\end{equation}
\begin{equation}
F_0^a = D_0^a+\frac{1}{2}F_0^sA_0^s-\frac{1}{2}F_0^aA_0^a,
\label{eq:Landau_Parameters_a}
\end{equation}
where, $A_0^{s,a}=F_0^{s,a}/(1+F_0^{s,a})=N(0)a_0^{s,a}$. In the unitary limit, the Landau parameters take on the following values: $F_0^s = -0.5$ and $F_0^a \to +\infty$. These parameters capture the strong interactions and successfully explain various universal thermodynamic properties of the unitary Fermi gas \cite{Li_Jackiewicz_Bedell_2015,Giorgini_Pitaevskii_Stringari_2008}.

Following the approach of Landau Fermi-liquid theory \cite{Baym_Pethick}, with the local induced interaction model, we calculate the quasiparticle scattering amplitudes $W_{\text{f}}(\theta,\phi)$ and $W_{\text{n}}(\theta,\phi)$:
\begin{equation}
W_{\text{f}}(\theta,\phi)=\frac{1}{2}W_{\uparrow\downarrow}= \frac{1}{2}\frac{2\pi}{\hbar}|a_0^{\uparrow\downarrow}|^2 = \frac{1}{2}\frac{2\pi}{\hbar}\Big|\frac{\mathscr{T}_s(\mathbf{K},0)}{2}\Big|^2
\label{eq:Scattering_Amplitude_f}
\end{equation}
\begin{equation}
W_{\text{n}}(\theta,\phi)=\frac{1}{2}W_{\uparrow\downarrow}=\frac{1}{2}\frac{2\pi}{\hbar}|a_0^{\uparrow\downarrow}|^2
=\frac{1}{2}\frac{2\pi}{\hbar}\Big|\frac{-2A_0^a}{N(0)}\Big|^2
\label{eq:Scattering_Amplitude_n}
\end{equation}
where $A_0^a = 1$ in the unitary limit. The total scattering probability becomes
\begin{eqnarray}
&\langle W\rangle& \,= \langle W\rangle_{\rm{normal}}+\langle W\rangle_{\rm{fluctuations}} \nonumber
\\
&=& \frac{2\pi}{\hbar}\frac{2}{|N(0)|^2}\cdot 2\left(1-\frac{\sqrt{6}\pi}{4\gamma}\frac{T_c}{T_F}\right)|A_0^a|^2 \nonumber
\\
&+& \frac{2\pi}{\hbar}\frac{2}{|N(0)|^2}  \nonumber
\\
&\times& \left[ \frac{\frac{\sqrt{6}\pi T_c}{4\gamma T_F}}{\ln\frac{T}{T_c} \left[ \ln\frac{T}{T_c} +\big(\frac{\sqrt{6}\pi T_c}{4\gamma T_F}\big)^2 \big(11.2+0.28 \big(\frac{T_F}{T_c}\big)^2 \big)\right]} \right. \nonumber
\\
&+& \left. \frac{\tan^{-1} \big( \sqrt{\big(\frac{\sqrt{6}\pi T_c}{4\gamma T_F}\big)^2 \big(11.2+0.28 \big(\frac{T_F}{T_c}\big)^2 \big)} \big/ \sqrt{\ln\frac{T}{T_c}}\big)}{\big(\ln\frac{T}{T_c}\big)^{3/2} \sqrt{11.2+0.28 \big(\frac{T_F}{T_c}\big)^2}}\right]. \nonumber
\\
\label{eq:Total_Scattering}
\end{eqnarray}

To calculate the viscosity of the unitary Fermi gas within the Landau Fermi-liquid theory, we need the viscous lifetime $\tau_\eta$ in addition to the scattering probabilities eqn.(\ref{eq:Total_Scattering}). In the low temperature limit, the viscous lifetime, $\tau_\eta^0$, is \cite{Baym_Pethick}
\begin{eqnarray}
\tau_\eta^0 &=& \frac{0.205\times8\pi^4\hbar^6}{m^3\langle W\rangle(k_BT)^2}\nonumber\\
&=&0.205\tau
\label{eq:Viscous_Lifetime_0}
\end{eqnarray}
where the bare mass and the effective mass are the same since we're operating in a local model, $\tau$ without any index is the quasiparticle lifetime, and the factor of 0.205 is from the different angular average of the scattering amplitude in the unitary limit.  A finite temperature correction is added to $\tau_\eta^0$ to give \cite{Dy_Pethick_1969}
\begin{eqnarray}
\tau_\eta &=& \frac{\hbar}{k_B T_F}\left(\frac{T_F}{T}\right)^2\bigg(\frac{\hbar|N(0)|^2}{0.205\times16}\langle W\rangle-3\pi\zeta(3)\nonumber \\ &\times&[0.202(A_0^a)^3+0.164(A_0^a)^2]\frac{T}{T_F}\bigg)^{-1}.
\label{eq:Viscous_Lifetime_Full}
\end{eqnarray}
The viscosity is then given by: 
\begin{equation}
\eta = \left\{
\begin{array}{ll}
\frac{1}{5}n p_f v_f \tau_\eta, & T\ll T_F \\
\\
n k_{\scriptscriptstyle{B}} T\tau_\eta = 3.4n\hbar \left(\frac{T}{T_F}\right)^{3/2}, & T\gg T_F
\label{eq:Viscosity}
\end{array} \right.
\end{equation}
Eqn. (11) for $\eta$ is the standard Fermi Liquid result \cite{Baym_Pethick}. Eqn. (12) for $T \gg T_F$ can be interpreted as the classical viscosity which is found upon taking a thermal average of eqn. (11). The classical lifetime  $\tau\propto\frac{\hbar}{k_B T_F}\left(\frac{T}{T_F}\right)^{1/2}$ \cite{Bruun_2011} is found by fitting to data for the viscosity coefficient \cite{Cao_Thomas_2011} and given as $\tau_\eta \approx 3.4\frac{\hbar}{k_B T_F}\left(\frac{T}{T_F}\right)^{1/2}$.
A natural concern in our work thus far is our use of the Landau Kinetic equation (LKE) to calculate the viscosity is the short, tending to zero, quasiparticle lifetime. In fact, the validity of Fermi liquid theory close to the transition temperature is still an open question that's still under debate \cite{Mueller}. Typically, the formal derivation of the LKE and subsequent calculations don't allow for arbitrarily short quasiparticle lifetimes and one resorts to other methods, such as the Kubo formalism, to calculate transport quantities when the quasiparticle picture is insufficient. Bruun and Smith performed a calculation \cite{Bruun_Smith} and show that corrections to the LKE result are small compared to those using the Kubo formalism.  Additionally, the entropy from Ku et. al. \cite{Zwierlein}, shown in Fig.\ref{fig:entropy}, exhibits Fermi liquid like behavior above $T_c$. Therefore, in spite of other work that claims Fermi liquid theory isn't valid \cite{Gaebler,Sagi}, we justify our approach through the entropy data closely resembling that of a Fermi liquid as well as work done using other methods that yield transport coefficients that minimally differ from LKE results. To calculate the ratio $\eta/s$, we also need the entropy density of the unitary Fermi gas. According to Fermi liquid theory \cite{Baym_Pethick}, the low temperature entropy density is given by
\begin{equation}
s = \frac{\pi^2}{2} nk_{\scriptscriptstyle{B}} \left( \frac{T}{T_{\scriptscriptstyle{F}}} \right) \left[1-\frac{\pi^2}{10}B^s \left( \frac{T}{T_F} \right)^2 \ln\left( \frac{T}{T^*} \right) \right],  \qquad T \ll T_{\scriptscriptstyle{F}}
\label{eq:Full_LowT_entropy_density}
\end{equation}
where $T^*\sim v_Fq_c/k_B\ll T_F$ is a cutoff temperature \cite{Baym_Pethick} ($q_c$ is a cutoff momentum defined by $\left|p-p_F\right|\ll q_c\ll p_F$), $B^s = -\frac{1}{2}(4-\frac{\pi^2}{6})$ for a local Fermi liquid in the unitary limit, and the logarithmic term stands for the finite temperature correction to the low temperature result. In the high temperature limit, the entropy density takes the form of a classical Fermi gas \cite{Pathria}
\begin{equation}
s = nk_{\scriptscriptstyle{B}} \left\{ \frac{5}{2}-\ln\left( \frac{n\lambda^3}{g} \right) \right\}, \qquad T \gg T_{\scriptscriptstyle{F}}
\label{eq:Entropy_HighT}
\end{equation}
where $\lambda = h/\left(2\pi m^* k_{\scriptscriptstyle{B}}T\right)^{1/2}$ is the thermal wave length, and $g=2$ for a two component Fermi gases.
\begin{figure*}
	\centering
	\includegraphics[scale=5.5]{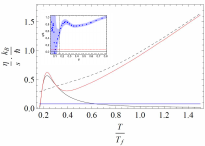}
	\caption{(COLOR ONLINE) The ratio $\eta/s$ vs temperature. The ratio $\eta/s$ is evaluated at $F_0^a = 100$, i.e. close to the unitary limit where $F_0^a \to +\infty$ according to the local model. The black solid curve is the low temperature limit of $\eta/s$ and the dashed curve is the high temperature limit. The red curve represents the single function that captures the behavior of both curves. The horizontal blue line indicates the quantum limited lower bound of $\eta/s=\hbar/4\pi k_{\scriptscriptstyle{B}}$ conjectured \cite{Kovtun_Son_Starinets_2005}. The inset figure shows the data for $\eta/s$ as a function of $\theta=T/T_F$ obtained by Joseph et. al. \cite{Joseph_Thomas_VS}. Their data seems to show $\eta/s=0$ at $\theta=0.1$, which agrees with our result if one considers that $T_c=0.1T_F$ in the local model. It should however be stressed that data in this region is inconclusive  and cannot be used to justify agreement with our result (For example, one can easily see that the ratio dropping to zero in the inset clearly happens well within the superfluid phase). However, what we can say about the inset is that the general behavior of their data is in good qualitative agreement with our result, albeit with a higher local minimum.}
	\label{fig:Viscosity/Entropy}
\end{figure*}
\begin{figure}
	\centering
	\includegraphics[scale=0.75]{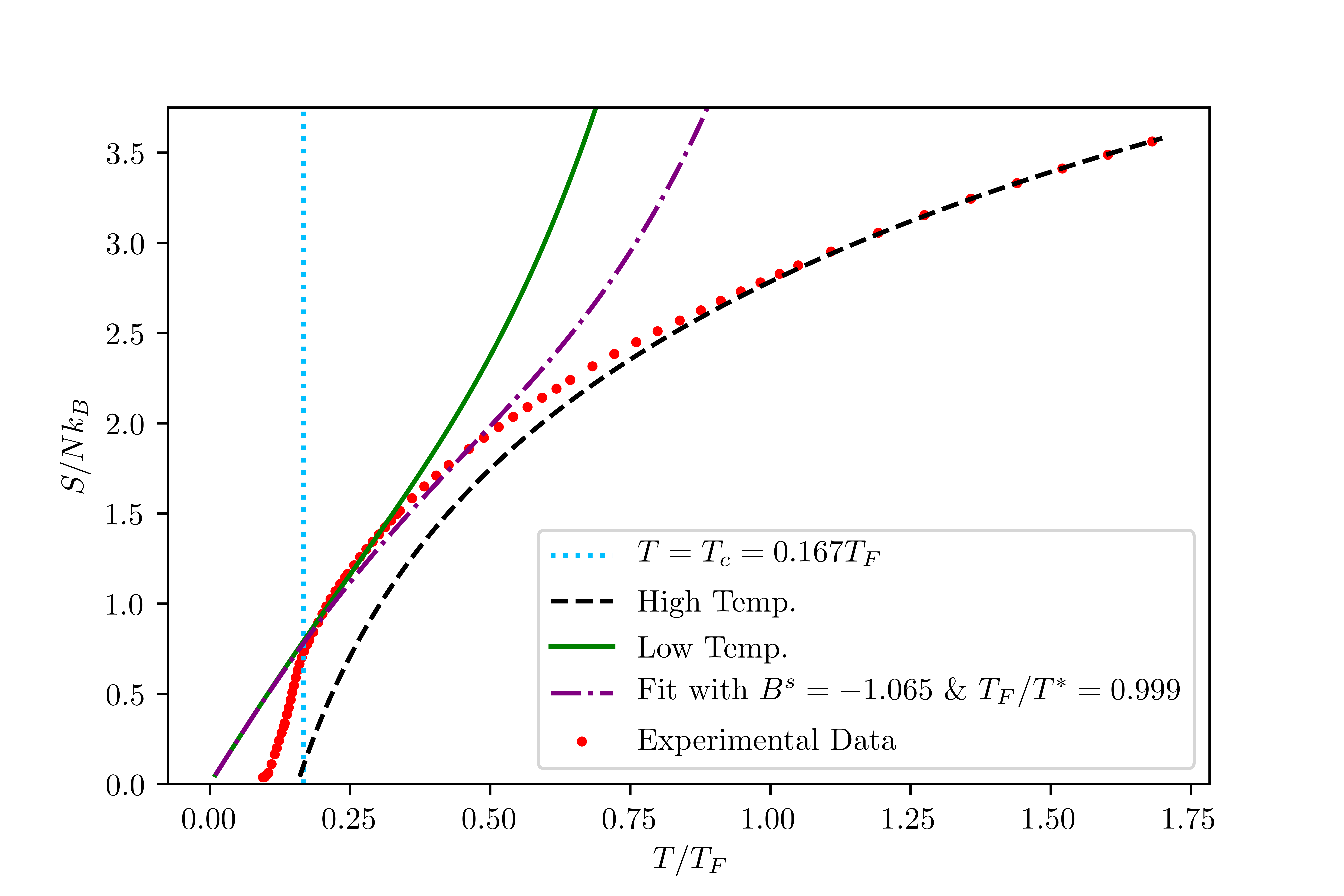}
	\caption{(COLOR ONLINE) Data for the entropy per particle from Ku et. al \cite{Zwierlein} (red dots). At all temperatures, specifically around $T_c=0.167T_F$ (blue dotted line), the entropy is a well behaved function without discontinuity. This supports the claim that $\eta/s\rightarrow0$ as $T\rightarrow T_c$ is due to lifetime effects and not unusual behavior in the entropy. The remaining three curves are our expressions for the entropy(eqn.(\ref{eq:Full_LowT_entropy_density}) and eqn.(\ref{eq:Entropy_HighT})). The green solid curve is eqn.(\ref{eq:Full_LowT_entropy_density}), the black dashed curve is eqn.(\ref{eq:Entropy_HighT}) and the purple dashed/dotted curve is eqn.(\ref{eq:Full_LowT_entropy_density}) with $T_F/T^*$ and $B^s$ being adjustable parameters (fit values given in legend). The purple curve, in spite of agreeing with the low temperature dependence and matching eqn.(\ref{eq:Full_LowT_entropy_density}), suggests that for more accurate results, we must go beyond the local model for a Fermi liquid. As one can see, the entropy behaves closely to that of a Fermi liquid suggesting a good quasiparticle picture and further validating our use of the LKE regardless of the vanishing quasiparticle lifetime.}
	\label{fig:entropy}
\end{figure}

The ratio $\eta/s$ is plotted over the entire temperature regime in Fig.\ref{fig:Viscosity/Entropy}. The experimental data of $\eta/s$ from \cite{Joseph_Thomas_VS}, shown in the inset of Fig.\ref{fig:Viscosity/Entropy}, is measured with respect to reduced temperature $\theta=T/T_F$. Additional data in \cite{Luo_Thomas_2009} plot the ratio with respect to $E/E_F$. A ratio of $E/E_F=0.6$ corresponds roughly to a temperature ratio of $T/T_F=0.17$, therefore the low temperature portions of our calculated and the measured ratios of $\eta/s$ are plotted within the same temperature window. A local minimum, with value $\eta/s \approx 0.3\hbar/k_{\scriptscriptstyle{B}}$, is found in the calculated ratio $\eta/s$ at $T\approx 0.36T_F$ (shown by the red curve in Fig.\ref{fig:Viscosity/Entropy}) agrees roughly with the experimental saturation value of $\eta/s$ for a nearly perfect Fermi gas \cite{Thomas_2010,Cao_Thomas_2011} (in the inset of Fig.\ref{fig:Viscosity/Entropy}) and is not far from the holographic prediction \cite{Kovtun_Son_Starinets_2005} $(\eta/s)_{\scriptscriptstyle{\text{KSS}}} = \hbar/4\pi k_{\scriptscriptstyle{\text{B}}}\approx 0.08\hbar/k_{\scriptscriptstyle{B}}$. However, as can be seen from Fig. \ref{fig:Viscosity/Entropy} and eqns.(\ref{eq:Total_Scattering}) and (\ref{eq:Viscosity}), the ratio $\eta/s$ is not bounded by this local minimum as it appears to drop to zero at $T_c$ due to superfluid fluctuations as
\begin{equation}
\frac{\eta}{s}\sim\left(\frac{T}{T_F}\right)^{-3}\left(\ln\left(\frac{T}{T_c}\right)\right)^{3/2}
\end{equation}
which qualitatively agrees with the behavior in the inset of Fig.\ref{fig:Viscosity/Entropy}. The conjectured universal lower bound for  $\eta/s$ is therefore violated in our theory. A concern with our result is if hidden behavior of the entropy density, not captured by eqn.(\ref{eq:Full_LowT_entropy_density}), is causing $\eta/s\rightarrow 0$. While eqn.(\ref{eq:Full_LowT_entropy_density}) may not be the complete low temperature behavior, the data given by Fig.\ref{fig:entropy} suggests that although a kink is present, there is no divergence or singularity. Neither theoretical (eqn.(\ref{eq:Full_LowT_entropy_density})) nor experimental result diverges and therefore we believe the entropy density is well behaved and isn't driving the ratio to zero. It is important to note that recent reanalysis of the data in the inset of Fig.\ref{fig:Viscosity/Entropy} was done by Bluhm et. al. \cite{Bluhm}. They observe a minimum slightly above $T_c$, as many other works do, but unfortunately cannot comment on a minimum at/below $T_c$. We believe the lack of conclusive results near $T_c$ is due to the volatile behavior of the system in close vicinity to the critical temperature. The two competing phases make it difficult to obtain data and theoretical results, ours included, are model dependent. What we can say however is that there is a finite quasiparticle weight \cite{Sagi} which lends to some validity in our result. Violation of the conjectured bound on $\eta/s$ within our model begs the following question: why do superfluid fluctuations in the unitary Fermi gas violate the KSS bound?

\section{Viscoelasticity of the Unitary Fermi Gas}
Previous work \cite{Visco2,Visco} has led us to study the connection between the viscoelastic behavior of the unitary Fermi gas and $\eta/s$. Alberte et. al have shown that holographic solids, solid massive gravity black branes with nonzero graviton mass, violate the KSS bound \cite{Visco2}. Their work ultimately found that holographic solids with a non-zero bulk modulus, specifically finite shear modulus, violate the KSS bound, with strong evidence for extension to real solids. Our work aims to go a step further by presenting a system where experiment is possible, the unitary Fermi gas, that exhibits viscoelastic behavior and violates the KSS bound.

We must first ask if the viscoelastic model is suitable to describe the unitary Fermi gas, i.e. if the following conditions are met: (i) $c_0$, $c_1\gg v_F$ where $c_0$ and $c_1$ are the speeds of zero and first sound respectively and/or (ii) $l\rightarrow0$ as $T\rightarrow T_c$ where \textit{l} is the viscous mean free path. Although (i) is violated for the unitary Fermi gas since $-1<F_0^s<0$, (ii) is satisfied since the quasiparticle mean free path goes to zero as $T\rightarrow T_c$ and Cooper pairs form. Additionally, provided we are in a regime such that $\omega\tau\gg 1$, according to 
\cite{LandauElastic}, the fluid behaves as a solid with elastic response. 

We start with the general form for the stress tensor for a viscoelastic model, different from those found in \cite{Baym_Pethick,Visco,LandauElastic}:
\begin{equation}
-\Pi_{ij}=\sigma_{ij}-\zeta \text{u}_{ll}\delta_{ij}
\label{eq:tensor}
\end{equation}
where 
$$
\sigma_{ij}=p\delta_{ij}+2\mu\left(\text{u}_{ij}-\frac{1}{3}\text{u}_{ll}\delta_{ij}\right)
$$
is the stress tensor that shows the two modes (an elastic mode which is $p\delta_{ij}$ and a shear mode which is the remaining terms) and
$$
\text{u}_{ij}\simeq\frac{1}{2}\left(\frac{\partial u_i}{\partial  x_j}+\frac{\partial u_j}{\partial x_i}\right)
$$
is the strain tensor for small displacements and $u_i$ is the flow velocity. $\zeta$ is the bulk viscosity and may be ignored since $\zeta/\eta\sim T^4$ at low temperature for a Normal Fermi Liquid \cite{Baym_Pethick,BulkVisco}. In general, $\mu$ is the shear modulus which contains the viscous (viscosity) and elastic (elasticity) behavior (i.e. there are in general two modes $\mu_\perp$ and $\mu_\parallel$). Within the viscoelastic model, due to the short lifetime near $T_c$ in the unitary Fermi gas, we have $\omega\tau_\eta\ll 1$, $\eta\sim\tau\mu$ and elasticity is no different from viscosity. Using the LKE we get
\begin{equation}
\omega^2-c_1^2q^2=\frac{2}{15}\left(qv_F\right)^2\left(1+\frac{F_2^s}{5}\right)\frac{\nu_2}{\nu_0}
\label{eq:w}
\end{equation}
\begin{equation}
\frac{\nu_2}{\nu_0}\simeq 2\left[1+i\left(1+\frac{1}{5}F_2^s\right)/\left(\omega\tau_\eta\right)\right]^{-1}
\label{eq:nu}
\end{equation}
where the real and imaginary parts of (\ref{eq:w}) are analyzed separately. Letting $\omega=c\left(q-i\alpha\right)$, we obtain the following expression for the coefficient of sound attenuation \cite{Pethick} in the unitary Fermi gas
\begin{equation}
\alpha=\frac{2}{15}\left(\frac{v_F}{c}\right)^2q\left(\frac{\omega\tau_\eta\left(1+F_2^s/5\right)^2}{\left(\omega\tau_\eta\right)^2+\left(1+F_2^s/5\right)^2}\right)
\label{eq:attenuation}
\end{equation}
The real part of (\ref{eq:w}) gives
\begin{eqnarray}
\left(c^2-c_1^2\right)q^2-c^2\alpha^2&=&\frac{2}{15}\left(qv_F\right)^2\left(1+F_2^s/5\right)\nonumber
\\
&\times&\left(\frac{2(\omega\tau_\eta)^2}{\left(\omega\tau_\eta\right)^2+\left(1+F_2^s/5\right)^2}\right)
\label{eq:first sound}
\end{eqnarray}
Eqns.(\ref{eq:attenuation}) and (\ref{eq:first sound}) provide experimentally attainable quantities relating to the viscoelasticity of unitary Fermi gases. As the temperature of the unitary Fermi gas approaches $T_c$, two things happen: (i) $\alpha\rightarrow0$ and (ii) $c\simeq c_1$. From \cite{LandauFluid} we interpret $\alpha\rightarrow0$ as the penetration depth of $c_1$ being infinite. Additionally, if we impose the restrictions of the local model, as mentioned earlier when dealing with the unitary Fermi gas near $T_c$, Fermi liquid parameters $F_1^s$ and higher are zero but the behavior of $\alpha$ and $c_1$ remain unchanged. As the unitary Fermi gas approaches its transition temperature, the zero sound mode predicted by Landau Fermi Liquid Theory is over damped and not propagating. This leads to the first sound mode propagating through the entire system and is another indicator of viscoelatic behavior. Continuing with the Landau Kinetic equation, we can use conservation laws (momentum and number) to obtain a hydrodynamic equation of motion for the mass density
\begin{equation}
\frac{\partial^2\rho}{\partial t^2}-c_1^2\nabla^2\rho=\frac{4}{3}\frac{\eta}{\rho_0}\frac{\partial}{\partial t}\nabla^2\rho
\label{eq:wave}
\end{equation}
where if $\eta=0$, as our result suggests, we obtain 
\begin{equation}
\frac{\partial^2\rho}{\partial t^2}-c_1^2\nabla^2\rho=0
\label{eq:wave1}
\end{equation}
a standard wave equation for a sound wave propagating at velocity $c_1$ which is in agreement with our analysis and interpretation of eqn.(\ref{eq:first sound}).

\section{Summary}
Superfluid phase transitions appear to have significant effects on the ratio $\eta/s$. Our work, investigating such effects in the unitary Fermi gas, presents a violation of the conjectured KSS bound thus calling into question its proposed universality as well as the role of phase transitions on $\eta/s$. In general, strongly coupled systems often exhibit phase transitions leading us to wonder if similar conclusions could be drawn about other strongly correlated quantum fluids. For example, in dense nuclear matter produced in heavy ion collisions, the ratio is found to be obeyed albeit taking on a very small value of  $(\eta/s)_{\scriptscriptstyle{\text{KSS}}} \leq \eta/s \leq 2.5(\eta/s)_{\scriptscriptstyle{\text{KSS}}}$ \cite{Cremonini_2011}. Based on our model, one could argue that the small value of $\eta/s$ is related to fluctuations that arise from the strongly interacting quark gluon plasma (QGP) phase \cite{Chen_Nakano_2007}.  The transition temperature for the QGP phase is predicted from lattice QCD computations \cite{Karsch_Laermann_Peikert_2001} to be, $T_{\scriptscriptstyle{\text{QGP}}} \sim 170 MeV$, and from the experiments below this temperature, $\eta/s$ is close to the KSS bound \cite{Jacak_Steinberg_2010}. Therefore, we raise a general question: Is the minimum found in $\eta/s$ of the nearly perfect quantum fluid due to universal quantum behavior predicted by the AdS/CFT correspondence or is it a local minimum in the ratio $\eta/s$ caused by the interplay between correlated liquid effects that want the ratio to grow and the fluctuations of a nearby phase that want to drive them to zero at/near the phase transition? 

The model developed in this work, which differs from other work by taking into consideration amplitude fluctuations, aims to study the ratio $\eta/s$ in strongly correlated quantum fluids. While past calculations find a minimum that obeys (\ref{KSS}), such as those calculated by Wlaz\l{}owski et. al. \cite{Wlazlowski,Wlazlowski2}, our calculations have shown that fluctuations from the nearby superfluid phase can drive the ratio $\eta/s$ to very low values, even to zero at the phase boundary, thus violating the conjectured universal lower bound. More precise measurements of $\eta/s$ near the phase boundaries, in tighter temperature windows around $T_c$, are needed to establish validity of the KSS bound. Additionally, we expand on the connection between viscoelastic responses and violation of the conjectured bound as was first introduced by Alberte et. al. \cite{Visco2}. In our work and that done by Alberte et. al., two systems that can violate the KSS bound, the unitary Fermi gas and holographic solids, exhibit both viscous and elastic responses implying that complicated viscoelastic behavior, in addition to phase fluctuations, contribute to violation of the KSS bound. In conclusion, our theory provides an alternative and unique way of studying $\eta/s$ in a strongly correlated quantum fluid by considering the effects of pairing instabilities in the quasiparticle scattering amplitude. We hope this work sheds light on the rich connection between condensed matter and high energy problems through (bottom up) AdS/CFT.

\section*{Acknowledgements}
The authors M. Gochan,  H. Li, and K. Bedell would like to thank Joshuah Heath for his valuable discussions on AdS/CFT, John Thomas for data and figure use in the inset of Fig. 3, and Mark Ku for experimental data used in Fig. 4. This work is supported by the John H. Rourke Boston College endowment fund.

\section*{References}
\bibliographystyle{iopart-num}
\bibliography{ViscosityBoundViolation}

\end{document}